# Super Low Traction under EHD and Mixed Lubrication Regimes

Philippe Vergne

*Laboratoire de Mécanique des Contacts et des Solides - LaMCoS*
*UMR CNRS - INSA-Lyon n°5514, Bâtiment Jean d'Alembert*
*20 avenue Albert Einstein, F-69621 Villeurbanne, France*

## 1. Introduction

*1.1. Superlubricity, near frictionless sliding and super low traction*

After the pioneered experimental works on superlubricity by Martin *et al.* on $MoS_2$ [1], Hirano *et al.* on tungsten and silicon [2] and the further confirmation by Dienwiebel *et al.* on graphite [3], many groups around the word investigated the occurrence of near frictionless sliding contacts. This large mobilization of tribologists, material sciences specialists and physicists has lead to emerging solutions involving new materials and coatings, the most promising being carbon based like graphite, diamond, carbon composites or diamond-like-carbons. Some of them are currently used in practical applications.

However the introduction of a fluid between two contacting surfaces remains the traditional and still the most efficient way to prevent contact failures when the operating conditions generate high contact pressures, large thermal dissipation or when the presence of worn films or particles is prohibited.
In the field of lubrication, super low traction doesnot probably have the same significance compared to superlubricity of carbon based materials which gives friction coefficients lying within 5% (near atmospheric conditions) to almost 0.1% (under vacuum) whereas values encountered under classical "dry" conditions are almost always greater than 10%-20%. The situation is different especially in EHL: the highest friction coefficients are close to 10% when traction fluids are involved, i.e. fluids that have especially designed to transmit the highest friction, and they vary within 3-6% for the rest of lubricants. The range of variation is consequently very narrow and these typical values are really low compared to those obtained in dry contacts: as a consequence the gain expected from a super low traction regime (defined in section 2.2) in lubrication will be probably more limited, especially in the case of experiments conducted at the meso or macro scales. This weak perspective could be one explanation on the relatively low number of articles in recent literature dealing with lubricated superlubricity in the above conditions.

Nevertheless there is still strong interest in this topic and more generally in the fundamental understanding of friction between lubricated surfaces. Dowson and Ehret [4] have recalled that typical EHD films were about one micron thick when the first solutions to the elastohydrodynamic problem were proposed. But this situation has changed with time and nowadays EHD films are of nanometer rather than micrometer proportions. This has been possible thanks to numerous contributions - both experimental and numerical – on film thickness build-up mechanisms published during the last 20 years that improved our knowledge on very thin EHL films, the influence of surface features etc. Nevertheless, very few of these publications also deal with friction.

A second interest concerns industrial applications that are developed with increasing demands for higher energy efficiency, durability, and environmental compatibility. Since friction is one of the main sources of lost energy in mechanical elements, it becomes a matter of urgency to propose innovative solutions to control and to optimize this parameter. An intermediate step would be a better understanding of the friction mechanisms under lubricated conditions and a significant improvement of friction prediction.

*1.2. Chapter objectives and summary*

In this chapter we will report and discuss the experimental appearance of super low friction forces that occurred in EHL or in mixed lubricated applications, i.e. tribological situations far away from those prevailing during nano or micro tribotests or during lubricated wear experiments. It means that we simulated lubricated contacts as those existing in real life, involving engineering surfaces and materials, applying representative speeds and normal loads. This domain is also called conventional tribology.

The title of this contribution mentions both elastohydrodynamic and mixed lubrication regimes. Compared to dry conditions, we specifically focused on the lubricant response according to two directions:
− From the rheological point of view to ensure that its behavior could favor super low traction under full EHD separation.
− Based on the classical Striebeck diagrams that normally present a minimum friction in the EHL regime, to analyze the transition region between EHL and mixed lubrication where lubricated superlubricity could occur.

## 2. Traction versus super low traction

*2.1. Generalities on EHD traction*

Friction or traction in highly loaded lubricated contacts results from complex and coupled phenomena that are not yet totally understood. Formally one would have to account for two distinct contributions: rolling friction that comes from the inlet pressure rise and shearing friction that results from a velocity difference of the contacting surfaces. However both numerical and experimental previous works [5-6] showed that the contribution of the former term is generally negligible compared to the latter. Consequently we will consider in the following that traction only results from the lubricant strength to slip and/or spin motions occurring in the high pressure region of the conjunction.

Slip, small film thickness and high contact pressure contribute to generate very important shear rates and very high shear stresses, as viscosity strongly increases with pressure this effect being one of those that allow the elastohydrodynamic lubrication (EHL) mechanisms to take place. Rheological and thermal effects can occur simultaneously. Moreover some unusual contact features reported in the literature suggest the occurrence of interfacial or boundary effects [4].

Compared to solid or dry lubrication, friction coefficient under EHL regime varies over a quite reduced range, from few % (see figure 1) to a maximum rising 10 to 12% in the typical case of traction fluids (figure 2). The traction coefficient (=friction coefficient) is usually evaluated and plotted as a function of the slide to roll ratio (*SRR* see equation 1) defined by the ratio of the sliding velocity that generates lubricant shearing (and hence friction) to the mean entrainment velocity that is an essential parameter in separation build-up.

$$SRR = \Delta U / U_e \qquad (1)$$

where  $\Delta U = U_1 - U_2$ is the sliding velocity,
and  $U_e = (U_1 + U_2)/2$ is the mean entrainment velocity.

The question of rating "super low traction" compared to "common traction" is developed in the next section together with the main related experimental issues.

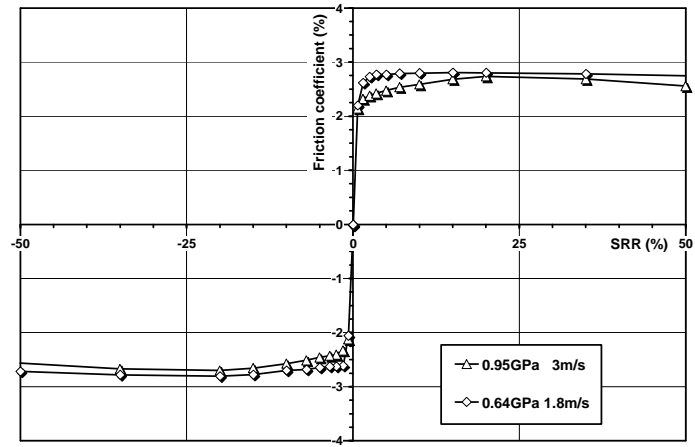

Figure 1: Low traction values given by linear paraffinic mineral base oil (LP) at 40°C.

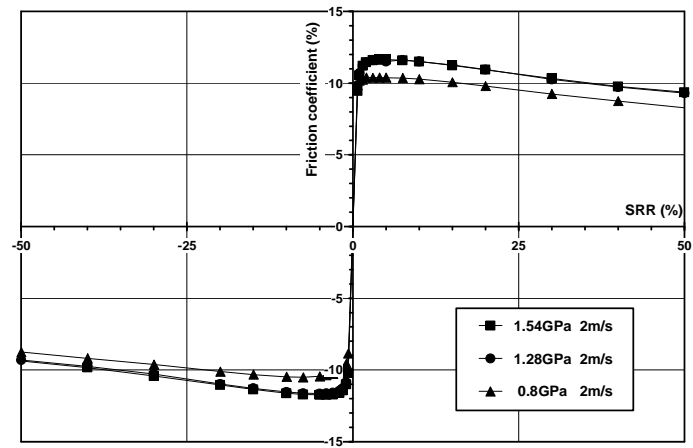

Figure 2: Typical high friction results obtained with a traction fluid (Santotrac 50, 25°C)

*2.2. Super low traction and experimental issues*

Compared to the two typical cases plotted in figures 1-2, super low traction characterizes not only situations where lower friction coefficients are encountered but also when different shapes of the traction curves are reported. A very steep friction increase from *SRR* = 0 is characteristic of lubricated contacts working under EHL conditions. Traction coefficients close to the maximum friction value are found for low slide to roll ratios, typically in the range 1 < *SRR* < 5%. However different behaviors can be found according to the operating conditions and the lubricant properties. Among them one can find the super low traction response that corresponds to lower friction values than those reported in figures 1-2. It can occur whatever the lubrication regime is, neither EHL nor mixed and gives a progressive friction increase with SRR. This peculiar behavior will be described and discussed in a next section.

Since under classical EHL conditions important friction values are obtained for even low *SRR* values, more attention has been paid in the past to control very accurately the velocities of the two specimens. This was also motivated by the existence of low *SRR* in ball bearings (typically few percents) that could generate much more traction variations than in the cases where *SRR* > 5%.
However, in the super low traction regime the experimentalist has to face another practical problem: the challenge is now to measure very low friction forces with adequate accuracy. This difficulty can be illustrated by the following practical example. For a given normal load - let's say 25N - it was classically relevant to assume a friction force sensibility of at least ±0.2% of the normal load (in our example ±0.05N). For the purpose of super low traction study, the requirement becomes more demanding and an acceptable sensibility should be 5 or 10 times higher than the value presented in the above example. In the super-low traction regime, this leads to sensibility lying in the range ±0.01 to ±0.005N, these values being considered as minimum ones. Facing this requirement, it becomes evident that numerous devices will be no longer adapted to study super low traction, as for instance conventional large twin disk machines or basic ball-on-disk set ups. This new demand has been already claimed by several authors in order to achieve significant breakthroughs in the understanding of superlubricity. Realistic devices have to be both well-controlled and relevant to operating machinery [7]. In their recommendations for future endeavors, Perry and Tysoe [8] mentioned that both nanoscale devices (like AFMs) and macro-tribometers need to be improved. They underlined the importance of technical points that are rarely discussed in papers: reproducibility, reliability, calibration, uncertainty analysis, sample and surface preparation, environmental control etc… They also pleaded for the improvement of the capacities of tribometers by a more precise control of forces and speeds.

### 3. Experimental conditions

Compared to micro scale or nanoscale investigations, we applied operating conditions closer to those found in real lubricated mechanisms like in rolling bearing elements or in automotive components: concentrated circular contacts, medium to high contact pressures, continuous motion of both specimens, variable slide to roll ratios, smooth surfaces, controlled lubricant feeding flow… These operating conditions were fulfilled by using a ball-on-disk test rig, similar to those designed to measure film thickness in EHD contacts and already described elsewhere [9-11].

A polished one inch ball of AISI 52100 bearing steel is loaded against a flat disk and both specimen are driven independently to allow for any desired slide to roll ratio. The ball and disk velocities are controlled with high precision and the cumulated geometrical defects are adjusted to minimize any fluctuation within the contact. The bottom of the ball dips into the reservoir containing the lubricant, ensuring fully flooded conditions. The contact and the lubricant are thermally isolated from the outside and heated (or cooled) by an external thermal controlling system. A platinum resistance probe monitors the lubricant temperature in the test reservoir within ± 0.1°C. Parts in contact with the lubricant are made from chemically inert alloys and any type of material likely to react with the fluid (rubber, elastomer) has been inhibited. The balls and the disks were carefully polished and cleaned following a three-solvent procedure to ensure minimum surface contamination.

Traction forces and normal load were recorded via a multi-axis strain gauge sensor. It combines a broad range of measurable forces, appropriate sensibilities over the different directions and high stiffness. This facility is directly positioned between the main frame of the test ring and the vertical assembly that includes the brushless motor, couplings, the shaft and its bearings and finally the disk. This design provides several important advantages:
− High linearity and sensibility due to a continuously applied prestressed state along the 3 directions.
− Only static parts are involved, leading to high signal to noise ratio compared to measuring systems attached to moving elements.

Several materials have been used for the disks: BK7 glass, pure synthetic sapphire and AISI 52100 bearing steel leading to composite RMS roughnesses of the undeformed surfaces of respectively 4, 5 and 9 nm. High pressure rheology and/or film thickness measurements have been carried out on most fluids investigated in this chapter. These previous experiments provided appropriate data for the evaluation of the $\lambda$ parameter as the lubrication regime will be an important factor in our analysis. $\lambda$ is defined as follows:

$$\lambda = h_{min} / \sigma \qquad (2)$$

where $h_{min}$ is the minimum film thickness,
$\sigma$ the composite rms roughness of the contacting surfaces.
Furthermore:
$\quad\quad\lambda > 3$ $\quad\quad$ gives full EHL separation,
$\quad\quad 3 > \lambda > 2$ $\quad\quad$ EHL regime but local contacts may occur,
$\quad\quad 2 > \lambda > 1$ $\quad\quad$ mixed lubrication regime,
$\quad\quad \lambda < 1$ $\quad\quad$ severe mixed lubrication regime.

**4. Lubricated super low traction**

This part reports and discusses experimental results where super low traction coefficients have been encountered. Several lubrication regimes (from EHL to mixed lubrication) and different types of lubricants will be considered. A table reported in annex summarizes the main rheological properties of these fluids together with their chemical composition and structure.
In the two first subsections, super low traction will be investigated under thick EHL conditions. Two really different lubricants will be studied and we will show that superlubricity can easily be explained thanks to the rheological behavior of each fluid.
The last subsection will concern mineral base oils, with and without additives. We will focus on the transition region between full EHL separation and mixed lubrication where film thickness becomes close to surface roughness. In this context special attention will be paid to the influence of the lubricants chemical structure.

*4.1. Newtonian isothermal piezoviscous behavior*

In this first section, the objective is to show how very low traction can be achieved using a simple viscous fluid that obeys in the simplest manner to EHD contact conditions. With this in mind we considered glycerol, a pure tri-alcohol characterized by its very compact molecular structure and low pressure viscosity coefficient. Results reported in figures 3-6 were obtained from various operating conditions varying the entrainment speed and the normal load and by changing the disk material from glass to sapphire and then to steel. Newtonian isothermal piezoviscous estimations of the friction coefficient based on the mean contact pressure $P_0$ (=$P_H.2/3$, $P_H$ being the Hertzian pressure) and Barus law are also plotted in these figures with dashed or full lines. Experimental results are represented by symbols only.
The traction response of this fluid is really different from that of the two boundary cases showed in section *2.1*, especially for the results plotted in figures 3-5. Firstly friction coefficient never exceeds 1% over the classical EHL range (0 < *SRR* < 50%) whatever the experimental conditions were. Secondly a linearly increasing friction coefficient is measured when the slide to roll ratio is increased. Finally a fair agreement is found between experimental results and those obtained from our basic numerical model.

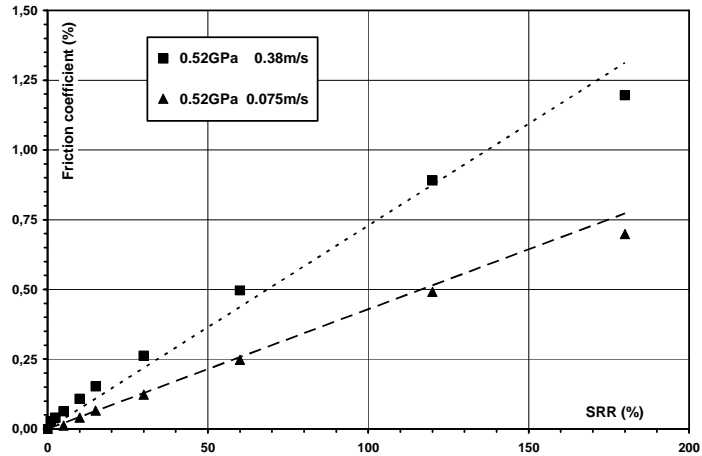

Figure 3: Traction curves measured with glycerol at 40°C, steel – glass contact.

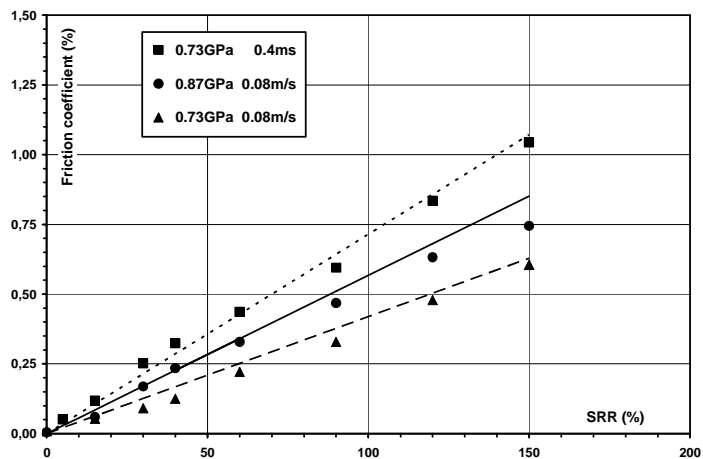

Figure 4: Traction curves measured with glycerol at 50°C, steel – sapphire contact.

However increasingly noticeable deviations appear when contact pressure and/or entrainment speed are increased, as it can be seen in figure 4 ($P_H$ = 0.87GPa, 0.08m/s), in figure 5 ($P_H$ = 0.95GPa) and in most cases plotted in figure 6.

Film thickness measurements – not reported here – showed that under pure rolling conditions (from 5mm/s to 5m/s, steel-glass contact, 40°C) glycerol behaved like a Newtonian piezoviscous fluid, as expected considering its simple and compact molecular structure. Shear thinning as may occur with polymers cannot occur with this fluid, only inlet shear heating has induced a drop on film thickness above 0.8m/s. These experiments also showed that high values of *SRR* had a weak influence on film thickness: thickness reduction of -7% at most for 0.38m/s, *SRR*=180% and 0.52GPa. This further study confirmed the idea that the applied power input and more specifically the pressure are likely the main causes of the observed deviations. A Hertzian pressure of 0.95GPa corresponds to a normal load that is 3 and 6 times the ones required to generate respectively 0.64GPa and 0.52GPa for a steel-steel contact. The term $\alpha P_0$ is sometimes used in the EHL community to describe the lubricant response: with glycerol, the deviation from the Newtonian isothermal piezoviscous behavior occurs when $\alpha P_0$ becomes greater than 3.1.

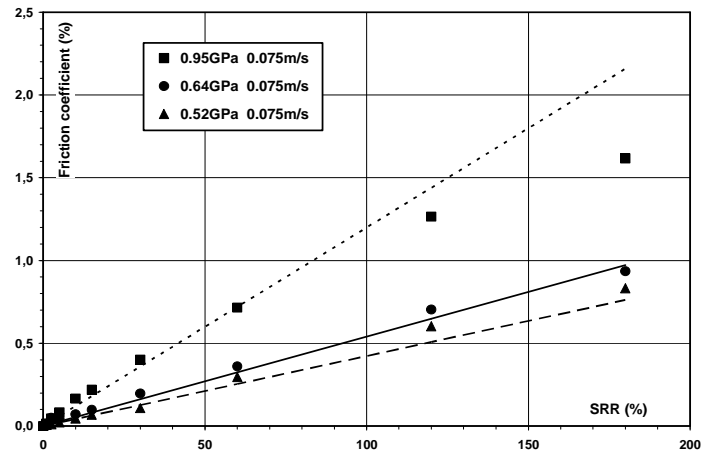

Figure 5: Traction curves measured with glycerol at 40°C and 0.075m/s, steel – steel contact.

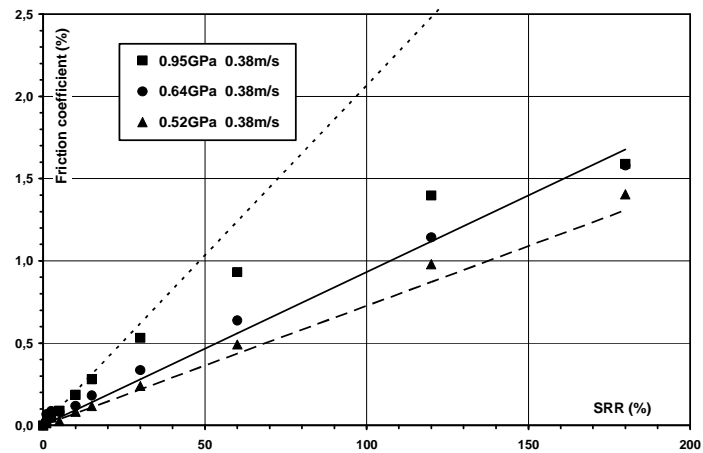

Figure 6: Traction curves measured with glycerol at 40°C and 0.38m/s, steel – steel contact.

Increasing the entrainment velocity also favors the deviation from Newtonian piezoviscous isothermal behavior. The transition occurs when *SRR* exceeds 60% at low speed (figure 5) and above 15% at 0.38m/s (figure 6). It is thus possible to estimate the corresponding in-contact mean shear rate by calculating the ratio of the sliding velocity to the central film thickness. This leads to values of $4.10^{+5}$ and $2.10^{+5} s^{-1}$ at respectively 0.075 and 0.38m/s; in rheological terms it means that the deviation from Newtonian isothermal behavior appears more rapidly when the entrainment velocity increases.

However one has to keep in mind that under the operating conditions of figures 5 and 6, glycerol also shows super low traction behavior. Measured friction coefficients remain lower than 1% when SRR varies from 0 to 50%.

*4.2. Shear thinning*

Shear thinning is probably the simplest and most frequent rheological effect that can reduce both film thickness and friction in an EHD conjunction lubricated by a conventional fluid. Shear thinning here concerns a steady shear rate dependence of viscosity. However it has been frequently combined with other effects like thermal heating and/or non linear high shear stress behavior, leading to certain confusion between the actual mechanisms that might influence friction.

Compared to the glycerol results reported in this section were obtained with a very different fluid, in terms of molecular structure, rheological properties and tribological response. The main objective in studying this lubricant model was to demonstrate that shear thinning that affects both film thickness and traction can be described by a unique ordinary non Newtonian relationship of the power law type [12]. This was accomplished by accurate measurements in viscometers (carried out by S. Bair at Georgia Tech) of the shear response of a high molecular weight polyalphaolefin (HMW PAO) and in-contact accurate measurements (performed at INSA de Lyon) of film thickness and traction under conditions which accentuate the shear thinning effect. This fluid possesses a very low critical stress for shear thinning, high viscosity at ambient temperature and pressure viscosity coefficient close to those of commonly used formulated lubricants. Moreover it also showed super low traction behavior.

In spite of these properties, experimental values of traction coefficient (black symbols in figure 7) are unusually low and sometime (at low speed and low slide-to-roll ratio) near the threshold of resolution of the sensor. In figure 7, the full line represents the Newtonian isothermal piezoviscous estimation of friction for the entrainment speed of 0.13m/s. Friction coefficients predicted by a numerical model assuming that shear thinning follows Carreau equation are plotted with dashed lines. Carreau equation is written:

$$\eta = \mu \left[ 1 + (\lambda . \dot{\gamma})^2 \right]^{(n-1)/2} \qquad (3)$$

where $\dot{\gamma}$ is the shear rate,

$\eta$ the generalized viscosity,
$\mu$ the low shear viscosity,
$n$ the power law exponent,
and $\lambda$ a characteristic time.

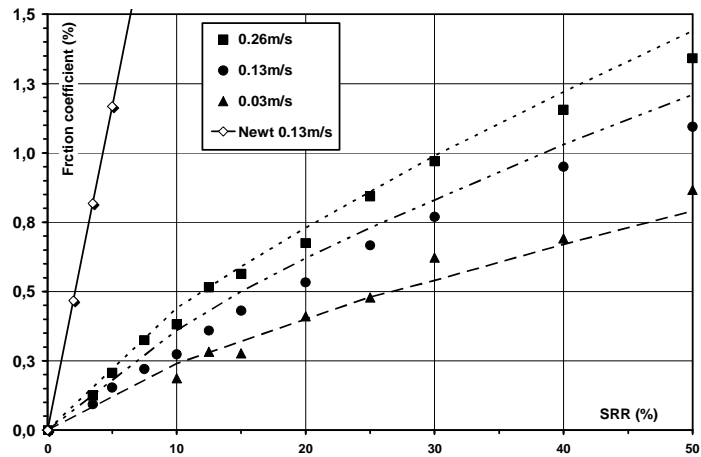

Figure 7: Super low traction obtained with a HMW PAO (75°C, 0.53GPa, steel-steel contact).

The Newtonian isothermal piezoviscous calculation greatly overestimates the actual traction behavior of this lubricant while the Carreau shear thinning model gives a good agreement with the measured data. We also showed [12] that central and minimum film thicknesses were insensitive to sliding and well described using the power-law relationship previously mentioned as equation (3). The in-contact central film thickness and traction were thus entirely predictable from the rheological properties obtained from viscometers using simple calculations. This proved that shear thinning - occurring mainly in the contact inlet - was the dominant effect that affected the shearing response of this fluid, in the absence of measurable thermal heating. As a consequence, the very low traction coefficients reported in figure 7 were attributed to this rheological behavior enhanced by the lubricant molecular nature.

*4.3. Thin film EHD conditions*

Results discussed so far were obtained under operating conditions that generated thick EHL films: in figures 3-7, the values of the λ parameter defined in equation (2) varied in the range 6-150.

In this section, we focus now on mineral base oils submitted to thin film tribological experiments performed under operating conditions where λ was lower than 7, and in most cases even lower than 3. All the tests were conducted on steel-steel contacts. Several objectives were followed: to check if these fluids studied under these specific conditions could give super low traction and to pursue investigations on the role of the lubricants molecular structure on their frictional behavior.

Mineral base oils of different structures were studied under 3 operating conditions: high contact pressure and high entrainment speed (0.95GPa, 3m/s, 40°C, figure 8), medium contact pressure but associated with lower entrainment speed at the same temperature (0.64GPa, 1.8m/s, 40°C, figure 9) and finally same pressure and speed conditions as in the second case but at higher temperature (0.64GPa, 1.8m/s, 70°C, see results in figure 10). These conditions lead to thinner and thinner film thicknesses and consequently they permit to increase the severity of the contact conditions

*4.3.1. Mineral base oils*

The nomenclature chosen for these fluids is detailed in annex: ARO, PN, ISO and LP present typical carbon chain length in the range $C_{11}$-$C_{14}$ as MIN contains longer chains ($C_{17}$-$C_{21}$).

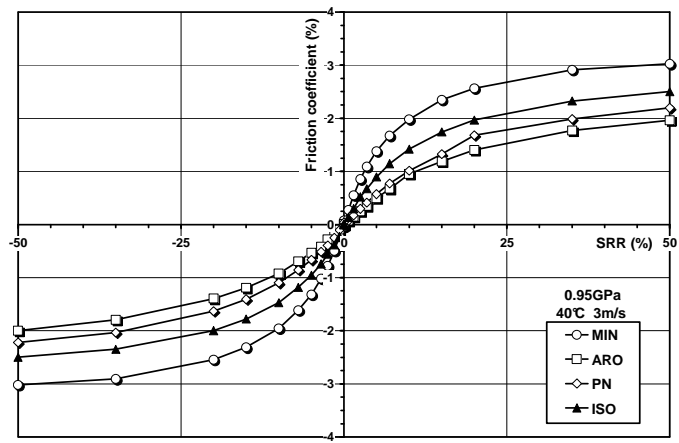

Figure 8: Traction curves obtained with different mineral base oils at 0.95GPa and 3m/s.

Except MIN and ISO, the lubricants exhibit similar rheological properties (see table in annex) leading to almost constant $\lambda$ values for given operating conditions. As a consequence, the results reported in figure 8 correspond to $\lambda$ equals to 7 (full separation) for MIN, 2.1 for ISO and 2.9 for the others fluids.
Apart from MIN that gives friction coefficients of 3% for SRR absolute values of 50%, ISO, PN and ARO fluids show moderate traction values and a progressive friction increase when SRR is increased. Base oil containing aromatic fractions (ARO) gave the lowest traction, then the mixture of paraffinic and naphtenic fractions (PN), the isoparaffinic base oil (ISO) and finally the more viscous fluid considered here (MIN).

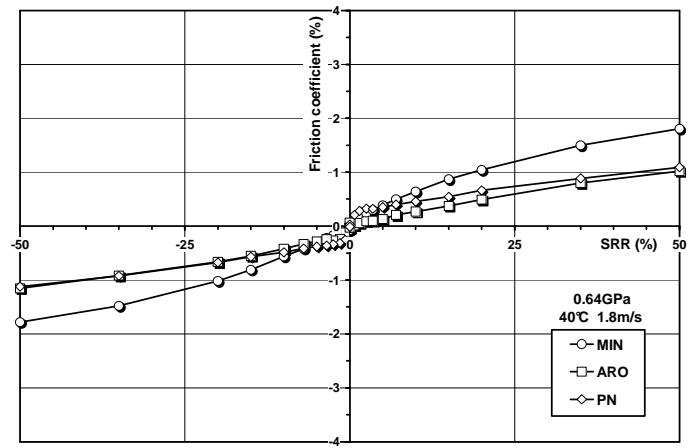

Figure 9: Low traction behavior obtained with 3 mineral base oils under more severe conditions than in figure 8.

Even if data reported in figure 8 show that both relatively low friction coefficients and smooth variations with SRR could be encountered, these preliminary results seem still far away from super low traction. The cases plotted in figure 9 represent $\lambda$ values of 5.4 for MIN and respectively 2.2 for PN and ARO: ISO traction results are not reported in this figure because this fluid (100% isoparaffinic) produced much higher traction coefficients (6 to 8%) than the other lubricants. This deviation is likely due to the inability of this base oil to maintain a low friction when operating conditions become more severe ($\lambda=1.7 \rightarrow$ mixed regime). This has been confirmed by further experiments where friction coefficients higher than 10% were found applying the same speed and normal load but increasing the temperature and consequently decreasing $\lambda$.
It should be noticed that under the operating conditions of figure 9, the low traction behavior of PN and ARO follows an almost linear increase when increasing the absolute value of the slide to roll ratio. A similar response has been found for the MIN base oil at 70°C where a maximum friction lower than 1% has been measured at $|SRR|$ = 50%, 0.64GPa and 1.8m/s.

LP traction results (see figure 1) have been obtained under the same operating conditions than those mentioned in figures 8-9. By comparison with base oils studied before and in spite of its low molecular weight, viscosity and pressure viscosity coefficient, the traction response of this 100% linear paraffinic fluid follows the classical EHL shape. Contact pressure has a dominant influence: rheological investigation carried out on this linear hydrocarbon showed the appearance of phase changes at very low hydrostatic pressures. Under the contact dynamic conditions all occurs as if the fluid was frozen within the conjunction and gave an almost constant friction coefficient only dependent on the slide to roll ratio sign. Moreover this behavior is consistent with solidification theories and visco-plastic models described in EHL literature. It also should be noticed that friction slightly increases when the pressure and the entrainment speed are decreased. This variation denotes a change from full EHL separation ($\lambda \approx 3$) to mixed lubrication regime ($\lambda \approx 2$).

These results showed that under mild contact pressures but under very thin film conditions ($2 < \lambda < 3$) super low traction can occur with simple mixtures of mineral fractions. This superlubricity regime should be considered as an optimum compromise that occurs over a quite narrow range of operating conditions. Further experiments run at higher temperature have confirmed that friction may significantly increase when $\lambda$ is approaching 1, according the lubricants chemical structure. It is not possible to advance a physical explanation on the near frictionless behavior of the mixtures, probably a favorable compromise between the paraffinic parts chain length and the lubricity contribution of naphtenic and aromatic fractions. However it is easier to understand why pure linear paraffinic (pressure induced rheological effect) and isoparaffinic (low viscosity and piezoviscosity coefficient, lower coverage of rubbing surfaces) base oils are unable to produce such interesting friction properties.

*4.3.2. Additive influence*
Results presented in the previous section were mainly obtained when $2 < \lambda < 3$, i.e. when local contacts between the specimen surfaces may occur that corresponds to the transition between full EHL separation and mixed lubrication. In this situation additives are usually introduced to extend the acceptable working range towards real mixed regime. Here we chose to add 8% w/w of lauric alcohol (noted LA in figure 10), actually a mixture composed of almost 70% of dodecanol and 30% of tetradecanol. This additive is considered as a lubricity improver (= friction reducer) and is used especially in the field of metal forming.

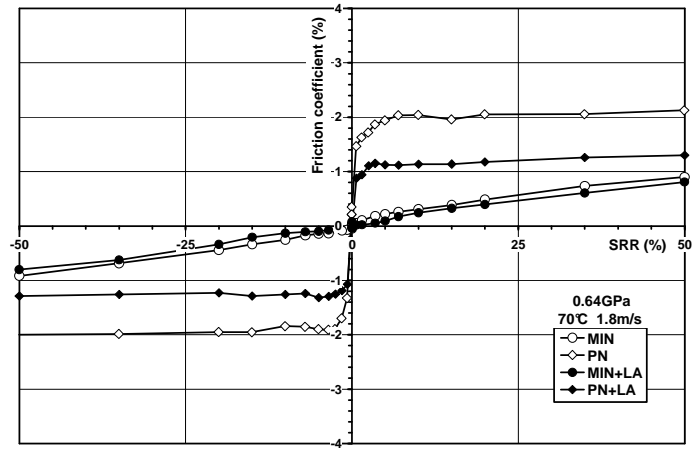

Figure 10: Comparison of traction coefficients obtained on base oils with and without lauric alcohol at 70°C.

Results will be discussed according the chemical structure of the base oils.
In presence of lauric alcohol, friction given by MIN is slightly lower than the one measured on the neat base oil whatever the operating conditions (see figure 10, 70°C). The additive has in this peculiar case (full separation even at 70°C) a rheological effect in reducing a little the film thickness. At 70°C a super low traction regime is achieved in both cases, with and without lauric alcohol.
Friction generated by ISO is considerably reduced when LA is added, however the traction coefficients remain higher than those found with the other lubricants and above the typical yield value that defines super low traction.
Even if the additive seemed to have a negligible influence on ARO friction at 40°C, the results obtained at 70°C show a significant but however limited increase (from 1.7 to 2.7%). The presence of aromatic fractions in the base oil suggests that an induced antagonistic effect between lauric alcohol molecules and the aromatic chains is the cause of the friction increase.
The tendency found with PN is totally opposed to the last one (ARO). Without additive we observe a friction increase when the temperature varies from 40 (figure 9) to 70°C (Figure 10), due to transition from soft to more severe mixed regime ($\lambda$ varies from 2.2 to 1.4). Traction drops when lauric alcohol is added. As for MIN+AL, one can consider that a super low traction regime is reached when PN+AL is used at this temperature.
Lauric alcohol addition permits to reduce NP friction for the three imposed operating conditions, but in any case values representative of super low traction regime were measured.

Under these thin film conditions, it has been shown that a super low traction regime may occur. The apparition of such lubrication regime depends on the lubricant structure and rheological properties. For different reasons pure isoparaffinic and pure linear paraffinic fractions gave relatively high traction coefficients whereas mixtures of paraffinic and naphtenic chains showed superlubricity, enhanced when fatty alcohols were added. Furthermore it should be underlined that operating conditions that permit to observe super low traction coefficients correspond to λ values lying in the range 1-3. This means that in some cases the classical Striebeck curve that usually presents a minimum friction value that coincide with EHL regime could be modified to take into account an almost frictionless behavior occurring at the transition between full film (EHL) and mixed lubrication regimes.

**5. Discussion and conclusion**

When trying to understand in detail the tribological mechanisms and interactions occurring in real contacts it would be useful to conduct separate analyses on three different scales, the macro, micro and nano scale and to study separately the mechanical and the tribochemical changes taking place in the contact. However much of the current information is fragmented, with linkages between individual results need have yet to be established.

In our case, results and subsequent analyses were based on macro scale tribological simulations. As an illustration, we showed how the rheological behavior can influence friction generated by the different fluids. Moreover, we may consider that the micro scale is of minor interest mainly because very smooth surfaces have been used. The nano scale analysis appears much more promising. The results from Krim [13] for instance, showed that liquid layers being more flexible and therefore slightly more commensurate with the surface, exhibited higher friction than their solid counterparts. However this finding has to be considered as a boundary case of our simulations for two reasons: it entailed liquid monolayers confined between extremely smooth and clean substrates made of high purity materials whereas our experiments involved at least several tens molecular layers sliding between bearing steel surfaces.

With the objective to merge information gained on the nano scale with that observed at the macroscopic scale, nonequilibrium molecular dynamic (NEMD) simulations may appear relevant. Jabbarzadeh *et al.* [14] showed the correlation between the degree of branching of $C_{30}$ alkane isomers and many important flow properties. They simulated Couette shear flow of thin (≈7nm) films submitted to very high shear rates. However, in spite of a great interest in the knowledge of lubricants high shear rate behavior, they assumed constant thickness and ambient pressure conditions. Bair *et al.* [15] proposed an approach that combined high shear rate NEMD simulations and high pressure rheological experiments to calculate EHD traction forces based on a Carreau shear thinning relationship. The comparison between simulated and experimental traction was successful but validated for only squalane. More recently Jabbarzadeh *et al.* [16] found that dodecane could exhibit a very low friction state when confined between perfect surfaces. However it appears that a confusion between shearing and friction exists and somewhat limits the contribution of this approach and the nanorheology one to a better knowledge of mechanisms that occur in lubricated macro contacts.

Another direction to make substantial progress in the understanding of friction and super low friction concerns the influence of the fluid structure. In [14], it is reported that linear alkanes should give lower friction (apparent viscosity), higher layering (and shear thinning) and lower slip than branched fluids of the same carbon chain length. Based on molecular interaction considerations and with the objective of molecular design of efficient traction fluids, some authors [17] showed that the lowest traction coefficients were obtained with fluids that could not interlock each other when they passed through the contact, like aromatic compounds. These tendencies qualitatively confirm some of our experimental results obtained on mineral base oils. However some weakness in the arguments (effect of contact pressure, real lubricant composition, engineering surfaces influence…) do not allow a formal link between nano and macro approaches.

To conclude on super low traction, it is expected that this concept will generate a similar scientific passion in the tribology community than thin film lubrication did in the early 90's. This desire is at first justified as friction reduction remains a major challenge to reduce energy losses, to improve durability of manufacturing goods and to meet constantly renewed environmental requirements. Another motivation for this lies in the fact that much more numerous applications will operate under conditions where lubricated superlubricity could occur. The use of low viscosity lubricants is one but an example: low viscosity could be induced by increasingly working temperatures (like in automotive engines) or an intrinsic property of the fluids like in applications where fuels have to lubricate mechanisms.

Concerning bridging the gap between nano and macro scale analyses and the improvement of friction understanding, it is expected that these issues can also contribute to a more quantitative prediction of traction coefficients. Obviously super low traction is included in these perspectives which could lead to extremely efficient solutions like for instance the one recently described by Kano *et al.* [18].

**Acknowledgements**

The author would like to thank several colleagues who accepted to share experimental data (S. Bair, Georgia Tech, Atlanta, USA) or who contributed to measurements on mineral base oils (F. Wiltord, former PhD student at LaMCoS) and on glycerol (K. Yagi, post-doctoral fellow at LaMCoS).
Special acknowledgments to J.-M. Martin (LTDS, Ecole Centrale de Lyon, France) who inspired the study on fatty alcohols and to Guillermo Morales (SKF-ERC and visiting professor at LaMCoS) for his valuable comments on the manuscript.

## ANNEX: Main properties of the lubricants

| Name | Type | Content | Structure | Viscosity (Pa.s) | $\alpha^*$ (GPa$^{-1}$) |
|---|---|---|---|---|---|
| Santotrac 50 | Traction fluid | Dicyclohexyl alkane + additives | $C_{18}H_{34}$ | 0.056 at 25°C [20] | 36 at 25°C [20] |
| Glycerol | Trialcohol | | $C_3H_8O_3$ | 0.293 at 40°C | 5.4 at 40°C [20] |
| HMW PAO | Synthetic hydrocarbon | Polyalpha olefine | M≈30000 kg/kmole | 1.42 at 75°C | 14.8 at 75°C |
| MIN | Mineral base oil | paraffinic + naphtenic | $C_{17}$-$C_{21}$ | 0.0047 at 40°C | 13.5 at 40°C |
| ARO | Mineral base oil | paraffinic + naphtenic + aromatic | $C_{12}$-$C_{14}$ | 0.0015 at 40°C | 10.2 at 40°C |
| PN | Mineral base oil | paraffinic + naphtenic | $C_{12}$-$C_{13}$ | 0.0015 at 40°C | 10.5 at 40°C |
| ISO | Mineral base oil | 100% iso-paraffinic | $C_{11}$-$C_{14}$ | 0.0011 at 40°C | 8.7 at 40°C |
| LP | Mineral base oil | 100% linear paraffinic | $C_{13}$-$C_{14}$ | 0.0015 at 40°C | 8.4 at 40°C |

Note that the pressure viscosity coefficient $\alpha*$ is actually the reciprocal asymptotic isoviscous pressure defined by Blok [19]. This parameter permits to account the pressure viscosity coefficient variation with pressure. When introduced in the classical EHL relationships, it gives the best agreement with measured film thicknesses.

**Additional references**